\documentclass[twocolumn,showpacs,amsmath,amssymb,prb]{revtex4}

\usepackage{graphicx}

\begin{document}

\title{Dynamical Decoupling of Qubits in Spin Bath under Periodic Quantum Control}
\author{Jun-Ting Kao$^{1}$, Jo-Tzu Hung$^{1}$,
Pochung Chen$^{1}$,  and Chung-Yu Mou$^{1,2,3}$} \affiliation{
1. Department of Physics, National Tsing Hua University, Hsinchu, Taiwan\\
2. Institute of Physics, Academia Sinica, Nankang, Taiwan\\
3. Physics Division, National Center for Theoretical Sciences,
P.O.Box 2-131, Hsinchu, Taiwan}

\date{\today}

\begin{abstract}
We investigate the feasibility for the preservation of coherence and
entanglement of one and two spin qubits coupled to an interacting
quantum spin-1/2 chain within the dynamical decoupling (DD) scheme.
The performance is examined by counting number of computing pulses
that can be applied periodically with period of $T$ before qubits
become decoherent, while identical decoupling pulse sequence is
applied within each cycle. By considering pulses with mixed
directions and finite width controlled by magnetic fields, it is
shown that pulse-width accumulation degrades the performance of
sequences with larger number of pulses and feasible magnetic fields
in practice restrict the consideration to sequences with number of
decoupling pulses being less than 10 within each cycle. Furthermore,
within each cycle $T$, exact nontrivial pulse sequences are found
for the first time to suppress the qubit-bath coupling to
$O(T^{N+1})$ progressively with minimum number of pulses being
$4,7,12$ for $N=1,2,3$. These sequences, when applied to all qubits,
are shown to preserve both the entanglement and coherence. Based on
time-dependent density matrix renormalization, our numerical results
show that for modest magnetic fields (10-40 Tesla) available in
laboratories, the overall performance is optimized when number of
pulses in each cycle is 4 or 7 with pulse directions be alternating
between x and z. Our results provide useful guides for the
preservation of coherence and entanglement of spin qubits in solid
state.
\end{abstract}


\maketitle

\section{Introduction}
The dream of building quantum computers has driven intensive
investigations on quantum information processing during the past
decade. Nonetheless, due to the ubiquitous decoherence problem, the
progress made so far has been limited. Since the processing of a
real quantum system includes inevitable disturbance from the outside
world, the central challenge is to find ways to control or even
eliminate the decoherence. There are different strategies proposed
to overcome decoherence such as dynamical decoupling (DD)
\cite{Lloyd1, Lloyd2,Viola,Lidar1,Lidar2, Das, Uhrig}, quantum error
correction\cite{Shor, Steane,Knill}, and decoherence-free
subspace\cite {Zanardi,Lidar,Facchi1,Facchi2}. While different
strategies have their own advantages, the dynamical decoupling
represents the oldest effort along this direction and have been
known as a mature technique employed in Nuclear Magnetic Resonance
(NMR) experiments. Theoretically, it has been rigorously shown that
DD provides upper bounds for error of reduced density matrix caused
by quantum evolution\cite{Lidar2}. Recent NMR experiments further
indicate that dynamical decoupling does preserve coherence of a
nuclear-spin qubit\cite{Lidar5}. These facts clearly indicate that
DD is promising in providing a practical solution to defeating
decoherence.

To implement the scheme of dynamical decoupling, explicit pulse
sequence has to be constructed.  Various pulse sequences were
proposed and developed. Hanh's spin echo (SE)\cite{Hahn} and
Carr-Purcell-Meiboom-Gill (CPMG)\cite {CPMG} were brought up in the
beginning. Later, concatenated dynamical decoupling (CDD)\cite{Das}
sequence and Uhrig's dynamical decoupling (UDD)\cite{Uhrig} sequence
were proposed. With so many pulse sequences available, one still
needs to address the central issue in the scheme of dynamical
decoupling: what is the sequence that has the best performance in
suppressing decoherence while viable quantum manipulations are kept?
The issue has been addressed by considering a given cycle of $T$ in
which pulses are applied. The performance is examined by number of
pulses needed for suppressing the qubit-bath coupling to the order
$O(T^m)$. When durations of pulses are ignored, it was recently
shown by Yang and Liu\cite{Liu} that for a single qubit interacting
with bath with Ising-like coupling, the UDD-N pulse sequence can
suppress the pure dephasing to $O(T^{N+1})$. However in addition to
the control of dephasing, one also needs to control longitudinal
relaxation. This would be necessary when the coupling between the
qubit and the spin bath is Heisenberg-like. In this case, Yang and
Liu\cite{Liu} showed that the UDD-N pulse sequence can not eliminate
the longitudinal relaxation and the dephasing to $O(T^{N+1})$ at the
same time. This calls for a closer examination on the minimum number
of pulses for suppressing the qubit-bath coupling to the desired
order $O(T^{N+1})$. Recently, a quadratic DD sequence (QDD) that
concatenating x-direction and z-direction UDD sequences is
proposed\cite{Lidar3}. Although QDD is shown to suppress general
decoherence to $O(T^{N+1})$ by using $(N+1)^2$ pulse intervals, the
sequence is not optimal and the issue of finding the optimal
sequence still remains.

From theoretical point of view, if the qubit-bath coupling is the
only Hamiltonian that governs qubits, the reduced density matrix
$\rho_r(T)$ of qubits includes all undesired dynamics. Therefore,
for a given $T$ and order $N$, a sequence is optimal if it
suppresses all operators in $\rho_r(T)$ to $O(T^{N+1})$. However, in
order to perform computing, one also needs a strategy for inserting
computing pulses\cite{Lloyd3}. It is clear that if the suppression
due to DD pulses is indiscriminating, the desired dynamics due to
computing will also be suppressed unless computing pulses form
another commuting DD-pulse sequence. In this case, one decomposes DD
pulses into cycles separating by computing pulses. The performance
of DD pulses is then examined by number of computing pulses that can
be applied before the system becomes decoherent. In addition to the
issue of how to insert computing pulses, the finite duration of
pulses also represents an important constraint. The accumulation of
pulse-width is seen to degrade the performance of DD
pulses\cite{Lidar5}. It is therefore important to compare
performance of sequences with different orders. So far, most
construction of DD pulses focuses on single qubit. It is known that
the entanglement is particularly important for characterizing the
quantum state of multi-qubits and plays the crucial role in quantum
information processing. There have been a few investigations of
effects of DD pulses on multiqubits. West et al. investigated
fidelity of quantum states of four nuclear spin-qubits in the
decoherence free space\cite{Lidar5} and found DD does preserve the
fidelity. There have also been studies based on pulse control of the
entanglement for two qubits in rather simplified models \cite{Abliz,
Rao, Masaki, Shan}. Nonetheless, it is still not clear what would be
an optimized sequence for preserving the entanglement.

In this paper, we investigate the feasibility for preservation of
decoherence and entanglement of spin qubits within the DD scheme in
solid state system. One or two spin qubits coupled to an interacting
quantum spin-1/2 chain are considered. We shall examine different
strategies for inserting computing pulses within each cycle and
demonstrate that computing after decoupling performs the best. We
then examine the feasibility by inserting computing pulses
periodically with period $T$ within which the same dynamical
decoupling pulses are applied. It is shown that error induced by
pulse-width accumulation restricts the consideration to sequences
with number of pulses being less than $10$ within each cycle.
Furthermore, within each cycle $T$, exact nontrivial pulse sequences
can be constructed to suppress the qubit-bath coupling to
$O(T^{N+1})$ progressively with number of pulses being $4,7,12$ for
$N=1,2,3$. Based on time-dependent density matrix renormalization
(t-DMRG), our numerical results show that for modest magnetic fields
(10-40 Tesla) available in laboratories, the overall performance is
optimized when number of pulses in each cycle is 4 or 7 with pulse
directions be alternating between x and z.

This paper is organized as follows. In Sec. II, we present our model
Hamiltonian and briefly discuss how to apply t-DMRG to analyze the
model Hamiltonian. We shall outline the general framework for
calculating the dephasing and longitudinal relaxation. In
particular, we point out that for general coupling between the qubit
and the bath, the preservation of either coherence or entanglement
is determined directly by the evolution operator $U(t)$. In Sec.
III, we analyze decoherence and longitudinal relaxation of a single
qubit by considering pulses with mixed directions. We will
explicitly construct pulse sequences for suppressing lower orders of
$U(T)$ up to $O(T^4)$. Furthermore, different strategies for
inserting computing pulses are compared. We find that computing
after decoupling performs the best. Therefore, we extend DD over a
cycle of $T$ to the periodic scheme in which computing pulses are
inserted at $nT$. By considering the finite duration of pulses, we
further analyze dynamics defined at $nT$. We show that for available
magnetic fields, the number of quantum manipulations can be
maximized by using sequence consisting of 4 or 7 pulses. Sec. IV is
devoted to investigate the entanglement of two qubits. We shall show
that regardless whether two qubits are strongly entangled or
non-entangled, the entanglement can be preserved by using the same
sequence that suppresses the decoherence of a single qubit. We
further show that for general multi-qubits scenario, entanglement
can be preserved by applying the same sequence to all qubits if
separations between qubits are sufficiently large. In Sec. V, we
summarize our results and discuss possible generalization to
dynamically decouple multi-qubits from the environment. In Appendix
A, we explicitly construct equivalent sequences for $N=2$.
\section{Theoretical Formulation and General Consideration}
\label{sec:model}

We consider a system-bath model which is described by the total
Hamiltonian $H_0=H_{sys}+H_{bath}+H_{int}$, where $H_{sys}$ is the
Hamiltonian of a single or two qubits system, $H_{bath}$ is the
Hamiltonian of a spin bath and $H_{int}$ represents the interaction
between qubits and the bath. The system Hamiltonian $H_{sys}$ is
generally zero unless computing pulses or decoupling pulses are
applied. Generally, computing pulses can be also spread over all
times\cite{Lidar5}. In this case, one has
\begin{equation}
\mathcal{H}_{sys} = \vec{s} \cdot \vec{B}, \label{uniformB}
\end{equation}
where $\vec{s}$ is the qubit spin operator and $\vec{B}$ is the
corresponding magnetic field for computing. The spin bath is a spin
chain generally characterized by the $XXZ$ Heisenberg model
\begin{equation}
\mathcal{H}_{bath} = J \sum \left( S^x_iS^x_{i+1}+S^y_iS^y_{i+1} +\Delta
S^z_iS^z_{i+1} \right),
\end{equation}
where $J > 0$. It is known that the XXZ Heisenberg model has a very
rich structure.\cite{Sachdev2000} The decoherence and entanglement
dynamics induced by such kind of spin bath have been recently
investigated. \cite {rossini:032333,Lai} In order not to be masked
by dynamics of the order parameter\cite{Hung}, we shall focus on the
XY regime where $|\Delta| <1$. Since the case of $\Delta \neq 0$
behaves qualitatively the same as $\Delta=0$ case, we shall simply
set $\Delta=0$ with understanding that our results are also
applicable to $\Delta \neq 0$. Two specific forms of the qubit-bath
coupling Hamiltonian $H_{int}$ are considered. In the first scenario
we consider Ising-like coupling $H_{int}=\epsilon s^z S^z_i$ for a
single qubit or $H_{int} =\epsilon ( s_1^z S^z_i+ s_2^z S_j^z)$ for
two qubits, where $s^z (S^z_i)$ is the qubit (spin bath) operators
and $i$($j$) is the single site of spin chain to which the qubit is
coupled to. It gives rise to pure dephasing of the qubit. To
minimize the boundary effect in numerical calculation, $i$($j$) is
usually taken to be the central site of the chain. In the second
scenario we consider Heisenberg-like coupling $H_{int}=\epsilon
\vec{s}\cdot
\vec{S}_i$ for a single qubit or $H_{int} =\epsilon ( \vec{s}_1 \cdot \vec{S}%
_i+ \vec{s}_2 \cdot \vec{S}_j)$ for two qubits, which gives rise to
both dephasing and energy relaxation.

\begin{figure}[h]
\includegraphics[width=8.5cm]{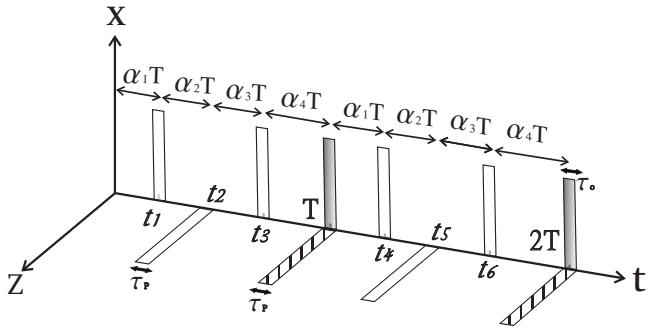} \caption{
Illustration of a dynamical decoupling pulse sequence in the
periodic decoupling scheme. Here a sequence of pulses, each with
width $\tau_p$ centered at $t_k$, are applied periodically with
period $T$. The directions of pulses are alternating and to
compensate the even/odd effect of pulse direction, a parity pulse
marked by slashed line is applied at the end of the sequence. At
$t^*=nT$, computing pulses of width $\tau_o$ for quantum processing
are applied. For convenience, we denote intervals between centers of
pulses by $\alpha_i T$.} \label{pulse}
\end{figure}

For most of our numerical work, we shall focus on the initial state
in which the total state of the system is a product state of the
form:
\begin{equation}
|\Phi(0)\rangle=|g \rangle \otimes |G\rangle, \label{eq1}
\end{equation}
where qubits are in some particular state of interest $|g \rangle$
while the bath is in its ground state $|G\rangle$. Our results,
however, are based on the consideration of evolution operator
directly. Hence similar results are also found for other initial
states.

A pulse sequence in a cycle of $T$ may contain $N$ decoupling pulses
centered at $t_k$ where $k=1,2,\cdots,N$ as illustrated in Fig.
\ref{pulse}. Because number of pulses along the same direction may
not be even, to compensate this parity effect, a parity pulse is
added at the end of the sequence. We shall denoted intervals between
$t_k$ by $\alpha_i T$. Typically in an experiment, these pulses are
generated by a magnetic field $B(t) \hat{n} $ and have the same
width $\tau_p$ . We concentrate on the pulse sequences in which each
pulse gives rise to a $\pi$ rotation along a given control axis
$\hat{n}$ for some qubit. If pulses are characterized by a
time-dependent control Hamiltonian $H_c(t)$, the total system is
then characterized by the Hamiltonian
\begin{equation}
H(t)=H_c(t)+H_0,
\end{equation}
with
\begin{equation}
H_c(t)=\sum_{i,k} A(t-t_k) \vec{s}_i \cdot \hat{n}^i_k.
\end{equation}
Here $\vec{s}_i$ represents the spin of $i^{\rm th}$ qubit and
$\hat{n}^i_k$ is the control axis during the $k^{\rm th}$ pulse that
is applied at the $i^{\rm th}$ qubit. $A(t-t_k)$ is a square
function centered at $t_k$ with width $\tau_p$. The magnitude of $A$
is $\mu_b B$ with $\mu_b$ being the Bohr magneton and $B$ be the
magnetic field so that $J \tau_p =10 \pi J ({\rm meV}) /B ({\rm
Tesla})$. Experimentally, accessible magnetic field $B$ will impose
a lower bound on the pulse duration $\tau_p$.

We shall focus on periodic dynamical decoupling where
$H_c(t+T)=H_c(t)$ and $T$ is the period of the pulse sequence.
Manipulating or computing pulses with width $\tau_o$ are applied at
$t^*=T, 2T, 3T, \cdots$. Since both $\tau_o$ and $\tau_p$ are
determined by available magnetic fields, we shall assume $\tau_o
=\tau_p$. Furthermore, in order that pulses are non-overlapping, one
requires
\begin{equation}
T > {\rm max}\left[ \frac{\tau_p}{\alpha_i} \right].
\label{nonoverlap}
\end{equation}
Since $\alpha_i ({\rm min})\leq 1/(N+1)$, we have $T \geq (N+1)
\tau_p$. Therefore, $T$ depends on $N$.
In the following we will denote $T$ by $T_N$ to indicate its explicit dependence on $N$.
Formally the evolution of the total system is dictated by the evolution operator
\begin{equation}
  U(t)=\mathcal{T} e^{-i\int_0^t H (s)ds },
\end{equation}
where we have set $\hbar$ to one and $\mathcal{T}$ is the
time-ordering operator. During each pulse period, the evolution
operator for the kth pulse can be written as
\begin{eqnarray}
  & & U_k \equiv \mathcal{T} exp^{ -i\int^{t_k + \tau_p/2 }_{t_k -
  \tau_p/2} H (s) ds } \nonumber \\
  &=&\mathcal{T} e^{ -i\int^{t_k}_{t_k-\tau_p/2} H_0 (s)ds }
  \mathcal{T} e^{-i \int^{t_k+\tau_p/2}_{t_k-\tau_p/2} H_c (s)ds} \nonumber \\
  & & \times  \mathcal{T} e^{ -i\int^{t_k+\tau_p/2}_{t_k} H_0 (s)ds }
  e^{ O(N \epsilon J \tau^2_p)} , \label{pusle}
\end{eqnarray}
where the second equality introduces an error of the order
$\tau^2_p$. Since for each qubit one has $\int^{t_k+\tau_p/2}_{t_k-\tau_p/2}
H_c ds = \pi \vec{s}_i \cdot \hat{n}^i_k$, the evolution operator
can be expressed as
\begin{eqnarray}
&& U^{(N)}(t) = (-i)^N e^{-i H_0 (t-t_N) } \Pi_i \vec{\sigma}_i
\cdot \hat{n}^i_N e^{-i H_0
(t_N-t_{N-1})} \times \nonumber \\
&& \Pi_i \vec{\sigma}_i \cdot \hat{n}^i_{N-1} e^{-i H_0
(t_{N-1}-t_{N-2})} \Pi_i \vec{\sigma}_i \cdot \hat{n}^i_{N-2}
\cdots e^{-i H_0 t_1} \times \nonumber \\
&& \left[ 1+ O(N \epsilon J \tau^2_p )\right], \label{UN}
\end{eqnarray}
where $\vec{\sigma}= 2\vec{s}$ are the Pauli matrices for the qubit
and the rotational matrix $e^{-i\pi(\vec{s}\cdot {\hat{n}}_k) }$ is
reduced to $i \vec{\sigma}\cdot {\hat{n}}_k$. Eq.(\ref{UN}) thus
implies that to the order of $\tau^2_p$, $N$ pulses of finite width
can be consider as ideal pulses without width
so that qubits are flipped right after $t_k$.

In general it is difficult to exactly evaluate $U^N(t)$. t-DMRG,
however, provides a way to efficiently evolve such a state with high
accuracy for a quasi-one dimensional system. We first use static
DMRG to find the ground state of the spin chain bath for a given
$\Delta$ and then use the method of t-DMRG to evaluate $U^N(t)|
\Phi(0)\rangle$ numerically. We note that the degrees of freedom of
the qubits are kept exactly during the t-DMRG calculation by
targeting an appropriate state. The dimension of the truncated
Hilbert space is set to be $D=100$. For short time simulation we set
$J\delta t=0.005$ in the Trotter slicing while for effective
dynamics we set $J\delta t=0.01-0.02$ to balance the Trotter error
and truncation error. Similar procedure has recently been used to
investigate the decoherence and entanglement dynamics induced by
spin bath. We hence refer to Ref.[\onlinecite{Lai}] and the
references therein for details of simulation procedure.

From $U(t)$, one obtains the reduced density matrix $\rho_r$ of qubits by
tracing out the environment
\begin{eqnarray}
  \rho_r(t) &=& Tr_{\text{bath}} {\ U(t) \rho (0) U^{\dagger} (t)}  \nonumber
  \\
  &=& \sum_E \langle \Phi_E |U(t) \rho (0) U^{\dagger} (t) | \Phi_E . \rangle
  \label{reduced1}
\end{eqnarray}
Here $|\Phi_E \rangle$ is a complete set of state for the spin chain and $%
\rho(0)$ is the total density matrix at $t=0$. Since the initial
total wavefunction is a product of state, one may consider $\rho (0)
= |p\rangle \langle q | \otimes |G\rangle \langle G |$, where $|p
\rangle$ ($|q \rangle$) is an eigenstate to the total $s^z$ of
qubits with the eigenvalue being $p$ ($q$). At time $t$, the reduced
density matrix is given by
\begin{equation}
  \rho_r(t) = \sum_E \langle \Phi_E | U(t) |p \rangle |G \rangle \langle G |
  \langle q| U^{\dagger} (t) | \Phi_E \rangle.
\end{equation}
For Ising-like coupling, $\hat{n}$ will be taken to be $\hat{x}$ so
that one can replace $\sigma^z$ by $\pm$. Therefore, one has
\begin{eqnarray}
U(t)|\pm \rangle & = &  e^{-i  (t-t_N)(H_0 \pm (-1)^N \epsilon S^z )
} \cdots e^{-i (t_2-t_1)(H_0 \mp \epsilon S^z ) } \nonumber \\
& & \times e^{-i t_1(H_0 \pm \epsilon S^z ) } |\pm \rangle  \nonumber \\
&\equiv & U_{\pm} (t) | \pm \rangle.
\end{eqnarray}
Hence one can replace $U(t) |p \rangle$ by $U_p(t) |p \rangle$. We
find
\begin{equation}
\rho_r(t) = |p \rangle \langle q| \left[ \sum_E \langle \Phi_E | U_p(t) |G
\rangle \langle G | U^{\dagger}_q (t) | \Phi_E \rangle \right] .
\end{equation}
Since $U_p(t)$ no longer acts on $|p \rangle$, one can switch the order of $%
\langle \Phi_E | U_p(t) |G \rangle$ and $\langle G | U^{\dagger}_q (t) |
\Phi_E \rangle$. Using the completeness of $|\Phi_E \rangle$, one obtains
\begin{equation}
\rho_r(t) = |p \rangle \langle q| \langle G | U^{\dagger}_q (t) U_p(t) |G
\rangle .  \label{reduced2}
\end{equation}
Hence the matrix element of the reduced density matrix is given by
\begin{equation}
\rho^{pq}_r(t) = \langle G | U^{\dagger}_q (t) U_p(t) |G \rangle .
\label{reduced3}
\end{equation}
It is clear that the effectiveness of DD control for the Ising-like
coupling is determined by $U^{\dagger}_q (t) U_p(t)$.

On the other hand, if the coupling between the qubit and the spin bath is
Heisenberg-like, $|p \rangle$ is no longer an eigenstate to $U(t)$. The
reduction from Eq.(\ref{reduced1}) to Eq.(\ref{reduced2}) is generally not
possible except for the diagonal elements\cite{Liu}. Therefore, one resorts
to Eq.(\ref{reduced1}) to calculate the reduced density matrix. In this
case, the matrix element of the reduced density matrix is given by
\begin{eqnarray}
\rho^{pq}_r (t) & = & Tr_{\text{bath}} \langle p | U(t) \rho (0) U^{\dagger}
(t) | q \rangle  \nonumber \\
&=& Tr_{\text{bath}} \left[ \rho (0) U^{\dagger} (t) | q \rangle
\langle p | U(t) \right]. \label{reducedpq}
\end{eqnarray}
In general, $| q \rangle \langle p |$ does not commute with $U(t)$.
Hence the effectiveness of bang-bang control for the Heisenberg-like
coupling is determined by the evolution operator $U(t)$. Note that
for longitudinal component when $p=q = \pm$ for a single
qubit, Yang and Liu\cite{Liu} noticed that for the UDD-N pulse
sequence applied at a single qubit, $U(T_N) = \exp \left[iH_{eff}
T_N + O(T_N^{N+1}) \right]$ and $H_{eff}$ commutes with $| + \rangle
\langle + |$ and $| - \rangle
\langle - |$. As a result, the linear term in $T_N$  gets canceled and thus the magnetization
$\langle \sigma_z (T) \rangle$ can be controlled to $1+ O(T_N^{N+1})$%
. Apparently, the same cancelation  does not happen for the off-diagonal
matrix elements where $H_{eff}$ does not commute with $|p \rangle
\langle q |$. Therefore, to find an effective sequence for both
dephasing and longitudinal relaxation, one needs to directly control
$U(t)$ to the required order.

\section{Short Time and Long time Dynamics of Single Qubit Decoherence}

In this section, we examine dynamics of a single qubit coupled to
the spin bath with and without decoupling pulses. For short time
dynamics of a qubit under DD pulses within a cycle of $T$, we
construct pulse sequences for suppressing lower orders of $U(T)$ up
to $O(T^4)$. Different strategies for inserting computing pulses are
compared. By using t-DMRG, we demonstrate that computing after
decoupling performs the best. Therefore, we extend DD over a cycle
of $T$ to the periodic scheme in which computing pulses are inserted
at $nT$. In particular, we shall compare long time dynamics of
different sequences at $nT$ to determine the optimized sequence.

\subsection{Ising-like coupling}

We start by examining the case when the qubit-bath coupling is
Ising-like. In this case, there is no longitudinal relaxation since
$[s_z, H]=0$. The pure dephasing is characterized by the Loschmidt
echo $L(t)\equiv\mid\rho^{+-}_{r}(t)\mid^2$, where $\rho^{+-}_r$ is
defined by Eq.(\ref{reduced3}). In the absence of decoupling pulses,
it is known that the Loschmidt echo decays as $L(t)=e^{-\alpha t^2}$
for short times\cite{Lai}. Hence $\alpha$ characterizes the short
time decoherence of a single qubit in spin
bath. When the spin bath is modeled by a XY model(i.e., $\Delta=0$) and $%
H_{int}=\epsilon s^z S^z_i$, $\alpha$ can be exactly calculated and will be
served as a checking point of our t-DMRG numerical code. To find $\alpha$,
we first note that after the Wigner-Jordan transformation, the Hamiltonian
is quadratic and is given by $H_0=J \sum_n c^{\dagger}_n c_{n+1} +h.c. +
\epsilon s_z c^{\dagger}_i c_i$ . As a result, $L(t)$ can be expressed as
\cite{Rossini}
\begin{equation}
L(t)=\mid\mathrm{det[1+r(e^{itH^-}e^{-itH^+}-1)]\mid^2.}
\end{equation}
Here $H^{\pm}$ are the matrices corresponding to $H_0$ with $s_z =
\pm 1$. $r$ is a $2N\times2N$ matrix with N is the length of the
spin chain and its element is given by $r_{ij}=\langle
c^{\dagger}_ic_j\rangle$. By using the identity for the operator
$\hat{A}$
\begin{eqnarray}
\det \left[1+ \hat{A} \right] &=& e^{Tr \ln \left( 1+ \hat{A}
\right) }
\nonumber \\
&=& e^{Tr \hat{A} - Tr \hat{A}^2 /2+ \cdots }
\end{eqnarray}
and expanding $e^{itH^{\pm}} = 1+ itH^{\pm} - t^2 (H^{\pm})^2 /2 \cdots$,
one find $\alpha = \epsilon^2$.

\begin{figure}[ht]
{\includegraphics[width=6cm]{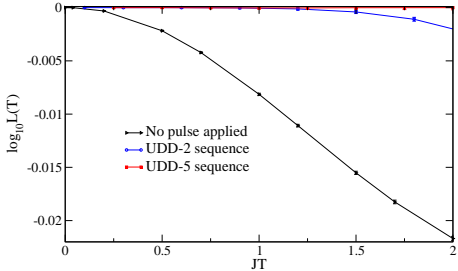}} \caption{{\protect\small
(Color online) Numerical simulations of Loschmidt echo $L(T)$ versus
$T$ by t-DMRG for Ising-like coupling. Here for a given $T$, all
pulses for any pulse sequence are arranged within $T$.  The fitted
decaying parameter $\alpha$ for the free case is $0.0224$ , in
agreement with the analytic result. It is clear that higher order
UDD pulse sequences are more effective. Note that in this and the
following figures, error bars are given for each data point with
figure legends being positioned at the intersection point of the
curve with the error bar. In the current figure, errors are about
the same sizes of figure legends.}} \label{IsingSingle}
\end{figure}

On the other hand, in the presence of N decoupling pulses in a
period $T_N$, it has been shown that the UDD-N pulse sequence suppresses\cite{Liu} $%
U_{-}^{\dagger(N)}(T_N)U^{(N)}_{+}(T_N)=1+ O (T_N^{N+1})$ and the
effectiveness increases as $N$ increases. As a check, in
Fig.\ref{IsingSingle}, we show our numerical simulations of $L(T)$
versus $T$ using t-DMRG. For the free decay without decoupling
pulses, fitting within $0<Jt<0.1$, we find that $\alpha=0.0224$ for
$J=1$ and $\epsilon=-0.15$. This is in agreement with the analytic
result $0.0225$ within the error caused by the Trotter time slicing
($J\delta t =0.005$). We also observe that higher order UDD pulse
sequence is more effective as expected.
\bigskip

\subsection{Short-time behavior of Heisenberg-like coupling}

When the coupling of the qubit to the spin bath is Heisenberg-like,
$s_z$ is no longer a good quantum number. To suppress both
longitudinal and transverse relaxations, we consider $N$
$\pi$-pulses with alternating directions along $x$ and $z$ axes in
the period of $T_N$. The evolution operator then becomes
\begin{eqnarray}
  && U^{(N)}(t) = (-i)^N e^{-i H_0 (t-t_N) } \sigma_x  e^{-i H_0
  (t_N-t_{N-1})} \sigma_z \times \nonumber \\
  & &  \cdots \sigma_z e^{-i H_0 (t_2-t_1)}\sigma_x  e^{-i H_0 t_1}
\left[ 1+ O(N \epsilon J \tau^2_p )\right]. \nonumber \\
\label{single}
\end{eqnarray}
Since for any operator $\hat{O}$, one has
\begin{equation}
  \sigma_{\alpha} \hat{O} (\vec{\sigma}) \sigma_{\alpha} = \hat{O}
  (\sigma_{\alpha} \vec{\sigma} \sigma_{\alpha}).
\end{equation}
By inserting appropriate identities, $\sigma^2_{\alpha}=1$, one can
move all the spin operators to the left and obtains
\begin{eqnarray}
  U^{(N)}(t) &=& (-i)^N \sigma_x \sigma_z \cdots \sigma_z \sigma_x U^{(N)}_0
  \left[ 1+ O(N \epsilon J \tau^2_p )\right] \nonumber \\
  &=& (-i)^{p} \sigma_p  U^{(N)}_0
  \left[ 1+ O(N \epsilon J \tau^2_p )\right]. \label{UN1}
\end{eqnarray}
Here $p$ is an integer and has no contribution in $\rho_r$.
$\sigma_p$ is a spin operator representing the net operation by
$\sigma_x \sigma_z \cdots \sigma_z \sigma_x$. For example, when
$N=3$, $\sigma_p = \sigma_z$. It is clear that $\sigma_p$ acts as a
parity pulse that compensate fast changes due to pulses. To remove
its effect, we must add an additional pulse $\sigma_p$ at $T$ so
that $U^{(N)}(t) = U^{(N)}_0 (t)$ where $U^{(N)}_0$ is the evolution
operator for ideal pulses without width and is given by
\begin{eqnarray}
& & U^{(N)}_0(t) = e^{-i H^{R_{\alpha_{N+1}}}_0 (t-t_N) } e^{-i
H^{R_{\alpha_{N}}}_0 (t_N-t_{N-1})} \times \nonumber \\
& & \cdots e^{-i H^{R_3}_0 (t_3-t_2)} e^{-i H^{R_2}_0 (t_2-t_1)}
e^{-i H^{R_1}_0 t_1}. \label{idealUN}
\end{eqnarray}

In this expression the time-dependent effective Hamiltonian is defined as
\begin{equation}
H^{R_n}_0 =H_{bath} +
\frac{\epsilon}{2}[ f_x(t) \sigma_x S^x_i + f_y(t) \sigma_y S^y_i + f_z(t) \sigma_z S^z_i ],
\end{equation}
with $f_\mu = \pm 1$, $\mu=x,y,z$ depending on the sequence. Consequently,
\begin{eqnarray}
& & U^{(N)}_0(t) = e^{-i H_{bath} t} \times \nonumber \\
& & \mathcal{T} e^{\left[ -i \frac{\epsilon}{2} \int_0^t f_x(s)
\sigma_x S^x(s) + f_y(s) \sigma_y S^y (s) + f_z(s) \sigma_z S^z (s)
ds
\right]} \nonumber  \\
& & \equiv  e^{-i H_{bath} t} G^N(t) . \label{timesequence}
\end{eqnarray}
Here $f_{x,y,z} (s)$ characterizes the sequence history of signs,
while $S^\mu (t)$ is the operator $S^\mu_i$ in the interaction
picture:
\begin{eqnarray}
  && S^\mu (t) = e^{iH_{bath} t} S^\mu_i e^{-iH_{bath} t}
  \nonumber \\
  && = \sum^{\infty}_{n=0} \frac{(it)^n}{n!}
  \underbrace{[H_{bath},[H_{bath},\cdots[H_{bath},S^\mu_i] \cdots
  ]]}_{n {\rm-folds}}
  \nonumber \\
  && \equiv  \sum^{\infty}_{n=0} S^\mu_n t^n . \label{expansion1}
\end{eqnarray}
It is clear that since $H_{bath}$ commutes with the qubit spin,
$e^{-i H_{bath}t}$ in Eq.(\ref{timesequence}) gets canceled in
Eq.(\ref{reducedpq}) Hence one only needs to suppress $G^N(T)$ to
the desired order. For this purpose, we re-express $G^N(T) =
\exp(\Omega^N(T))$ and use the Magnus expansion\cite{magnus}
$\Omega^N(T) = \Omega^N_1 + \Omega^N_2 + \cdots$ to evaluate
$G^N(T)$, where by setting $A(s)=f_x(s) \sigma_x S^x(s) + f_y(s)
\sigma_y S^y (s) + f_z(s) \sigma_z S^z (s)$, one has
\begin{eqnarray}
&& \Omega^N_1 (T) = \int^T_0 A(s) ds \label{Omega1} \\
&& \Omega^N_2 (T) = \frac{1}{2} \int^T_0 d s_1 \int^{s_1}_0 d s_2
[A(s_1), A(s_2)] \label{Omega2}.
\end{eqnarray}

If we want to suppress the decoherence to $m$-th order we must
suppress $U^{(N)}_0 (T_N)$ to $O(T_N^m)$, but at the same time due
to the finite width of pulses we also need to ensure $O(N \tau^2_p)
< O(T_N^m)$. Using Eq.(\ref{nonoverlap}), we find the minimum of
$T_N$ is determined by
\begin{eqnarray}
  T_N & \gtrsim & {\rm max}\left[ (N+1) \tau_p, (N J^{2-m}
  \tau_p^2)^{1/m}
  \right] \nonumber \\
  & \gtrsim & {\rm max}\left[ \frac{\tau_p}{\alpha_i}, (N J^{2-m}
  \tau_p^2)^{1/m} \right]. \label{minT}
\end{eqnarray}
We note that when $m$ is small, the first term dominates. It is thus
sensible to define a minimum period as $T_c= \tau_p/ min(\alpha_i)$.
Another important observation is that due to the finite width of
pulses, increasing number of pulses also increase $T_N$ which leads
to stronger decoherence. To find long-time dynamics at $t^*=nT_N$,
we shall start from $m=1$ and focus on small $m$. For a fixed $m$ we
find the minimum number of pulses needed, identify the optimized
value of $\alpha_i$, and compare the results from different $m$.
Indeed, as we shall see, increasing $m$ degrades the long-time
dynamics in some cases.

To suppress $ G^N (T_N)$ to $O(T_N^m)$, we keep the mth order term
in Eqs.(\ref{Omega1}) and (\ref{Omega2}). For $m=1$, there are three
independent operators, $\sigma_{\mu} S^{\mu}_i$ ($\mu =x,y,$ and
$z$), whose coefficients have to vanish. We hence obtain three
constraints
\begin{equation}
  \int^{T_N}_0 f_\mu (t) dt =0. \label{OT}
\end{equation}
These constraints can also be obtained from a geometric perspective,
as pointed out in Ref.\onlinecite{chen2006}. Therefore we find that
the minimum number of pulses is $N_1=4$ (including the parity
pulse). Eqs.(\ref{OT}) provide three equations for intervals
$\alpha_i$ where $i=1,2,3,4$. Together with $\sum_i \alpha_i =1$, we
find that there is only one solution with alternating $x$-$z$,
equally spaced pulse sequence
\begin{equation}
\alpha_1=\alpha_2=\alpha_3=\alpha_4=\frac{1}{4}. \label{solution1}
\end{equation}

For $m=2$, there are two additional contributions in
Eqs.(\ref{Omega1}) and (\ref{Omega2}) to $O(T_N^2)$ in $ G^N (T_N)$.
By using Eq.(\ref{OT}), terms with double integrals can be rewritten
as
\begin{equation}
  I_2 \equiv \int^{T_N}_0 dt_1 \int^{t_1}_0 d{t_2}
  f_{\mu} (t_1)  f_{\nu} (t_2) [S^{\mu}(t_2), S^{\nu}(t_1) ]. \label{double}
\end{equation}
Since $[\sigma_{\mu} S^{\mu}_i,\sigma_{\nu} S^{\nu}_i]=0$, all
double integrals vanish and only single integrals contribute. There
are three independent operators in Eq.(\ref{Omega1}). Consequently,
in addition to Eqs.(\ref{OT}), we require the 1st moment of $f_\mu$
to vanish
\begin{equation}
  \int^{T_N}_0 t f_{\mu} (t) dt =0. \label{OT2}
\end{equation}
By solving Eqs.(\ref{OT}) and (\ref{OT2}) for all possible
directions of pulses, we find that there are 5 solutions. To
continue from the case of $m=1$, we shall adopt the alternating
$x$-$z$ pulse sequence and refer the reader to the Appendix for the
remaining sequences. Hence, by including the parity pulse
($y$-pulse), the minimum number of pulses for $m=2$ is $N_2= 7$ with
intervals for six pulses being
\begin{eqnarray}
  & & \alpha_1= \frac{7-\sqrt{33}}{16}=\alpha_7, \hspace{0.2cm} \alpha_2=\frac{1}{8}=\alpha_6,
  \nonumber \\
  & & \alpha_3=\frac{\sqrt{33}-3}{16}=\alpha_5, \hspace{0.2cm}
  \alpha_4=\frac{1}{4}. \label{OT3sequence}
\end{eqnarray}
Similarly for $m=3$ the second moment of $f_\mu$ has to vanish.
\begin{equation}
\int^{T_N}_0 t^2 f_\mu (t) dt =0. \label{OT3}
\end{equation}
Additionally, there are six mixed moments which have to vanish too
\begin{eqnarray}
  \int^{T_N}_0 dt_1 \int^{t_1}_0 dt_2 (t_1-t_2) [ f_{\mu} (t_1)
  f_{\nu} (t_2) + f_{\mu}(t_2) f_{\nu} (t_1)] =0. \nonumber \\
  \label{OT4}
\end{eqnarray}
Therefore, in general, the minimum number of pulses for $m=3$ is
$N_3= 15$. However, for spin-$1/2$ qubits, since $\sigma^2_{\mu}=1$
and $(S^{\mu})^2=\hbar^2/4$, terms with $\mu=\nu$ in Eq.(\ref{OT4})
have no contribution to dynamics of qubits. Therefore, the minimum
number of pulses for a spin-$1/2$ qubit is $N_3= 12$. The equations
of $\alpha_k$ imposed by Eq.(\ref{OT4}) is generally very
complicated. Therefor, one has to resort to numerical methods to
obtain solutions. By solving Eqs.(\ref{OT}), (\ref{OT2}),
(\ref{OT3}) and (\ref{OT4}) for alternating $x$-$z$ pulse sequence,
we find the numerical values for intervals are
\begin{eqnarray}
&& \alpha_1 \cong 0.0171, \hspace{0.1cm} \alpha_2 \cong 0.0468,
\hspace{0.1cm}
\alpha_3 \cong 0.0658 \hspace{0.1cm} \alpha_4 \cong 0.1013, \nonumber \\
&& \alpha_5 \cong 0.1184, \hspace{0.1cm} \alpha_6 \cong 0.1006
\hspace{0.1cm}
\alpha_7 \cong 0.1195, \hspace{0.1cm} \alpha_8 \cong 0.1049, \nonumber \\
&& \alpha_9 \cong 0.0823 \hspace{0.1cm} \alpha_{10} = 0.1025
\hspace{0.1cm} \alpha_{11}\cong 0.0647 \hspace{0.1cm} \alpha_{12}
\cong 0.0439, \nonumber \\
&& \alpha_{13} \cong 0.0318. \label{solution2}
\end{eqnarray}
This is the optimized sequences for suppressing $U(T)$ to $O(T^4)$.
As indicated in the beginning, due to finite width of pulses, $m$-th
order (or $O(T^m)$) sequence is not necessarily more effective than
the $(m-1)$-th order (or $O(T^{m-1}))$ sequence. Hence we shall stop
at $O(T^4)$ and compare the performance of different orders at $nT$.

\begin{figure}[ht]
{\includegraphics[width=8cm]{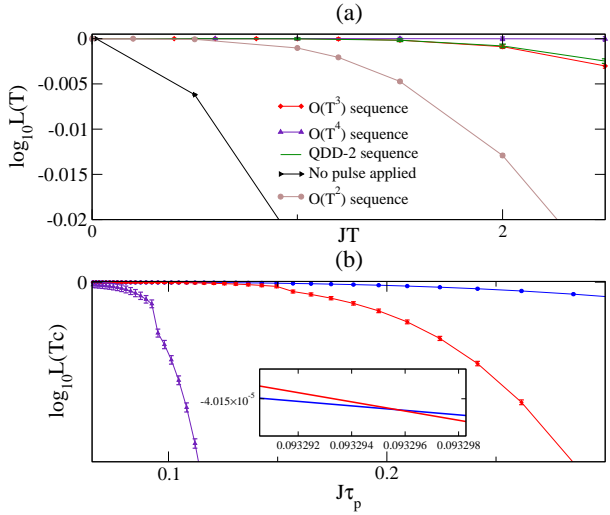}} \caption{{\protect\small
(Color online) (a) Loschmidt echo $L(T)$ versus $T$ of different
orders simulated by t-DMRG for Heisenberg-like coupling with
$\epsilon = -0.3$. Note that errors are about the same sizes of
figure legends. Clearly, $L(T)$ gets improved for increasing number
of orders. (b) $L(T_c)$ versus $J\tau_p$. It is seen for as $\tau_p$
decreases, $O(T^3)$ sequence starts to win over $O(T^2)$ sequence.}}
\label{HeisenbergSingle}
\end{figure}

In Fig. 3(a), we first show numerical results of $L(t)$ for $0 \leq
t \leq T$ based on t-DMRG for different sequences with a fixed $T$.
This corresponds to the ideal pulse scenario in which the pulse
width is neglected. Here we do observe that higher order sequences
are more effective. To take the finite width into consideration, in
Fig. 3(b) we show numerical results of $L(T_c)$ versus $J\tau_p$ for
sequence of different order. One should keep in mind that $T_c$
depended on $\tau_p$. One can see clearly in Fig. 3(b) that as
$\tau_p$ decreases, $O(T^3)$ sequence starts to win over $O(T^2)$
sequence. We note that for a much smaller $\tau_p$  the $O(T^4)$
sequence will out-perform $O(T^2)$ and $O(T^3)$ sequences (not shown
in the figure). As a check on if our results depend on the initial
state of the spin chain, in Fig. 4, we compare the Loschmit echo
versus $T$ by using $|\Phi(0)\rangle=|g \rangle \otimes |bath
\rangle$ with the state of bath $|bath \rangle $ being the ground
state or the first excited state of the quantum spin chain. One can
see that their difference is within the error bar. In addition, we
can also form entangled state of the qubit and the bath to check the
performance of our sequences. Since after the first cycle $T$, qubit
and bath is entangled. Therefore, the performance at later times is
an indirect check and this will be done in the next section.
\begin{figure}[ht]
{\includegraphics[width=6.5cm]{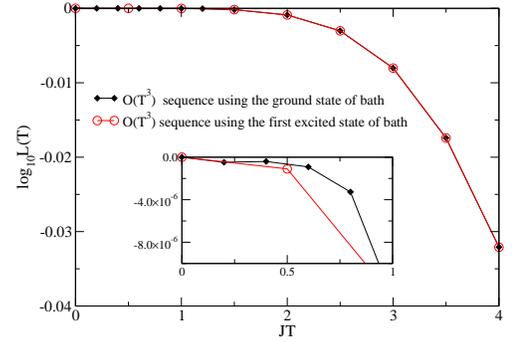}} \caption{{\protect\small
(Color online) Comparison of Loschmit echo by using different
initial states. Note that errors are smaller than sizes of figure
legends. }} \label{HeisenbergSingleR}
\end{figure}

In Fig. 5, we compare longitudinal relaxation under UDD-3 and our
optimized sequences. It is clear that the optimized sequences
outperforms the UDD-3 sequence.
\begin{figure}[ht]
{\includegraphics[width=8cm]{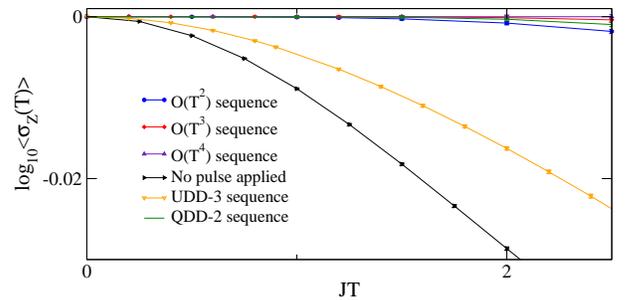}} \caption{{\protect\small
(Color online) Comparison of longitudinal relaxation of different
orders with UDD-3 and QDD-2 sequences. Here $\epsilon = -0.3$ and
errors are about the same sizes of figure legends. }}
\label{HeisenbergSingleR}
\end{figure}

\subsection{Long time dynamics of Heisenberg-like coupling}

Given the dynamics of dynamical decoupling, one needs a strategy for
inserting the computing pulses. For this purpose, we compare three
possible ways to insert a computing pulse that rotates the qubit by
an angle $\theta$: (i) using a constant $B$ over $T$ in
Eq.(\ref{uniformB}) (ii) applying a short pulse to rotate the qubit
by $\theta$ at some moment $t$ with $0<t<T$ (iii) applying a short
pulse at $T$ after all decoupling pulses. From the construction of
$O(T^N)$ sequences, since $H_{sys}$ in Eq.(\ref{uniformB}) can be
combined with $H_{int}$ with $\vec{S}_i$ being replaced by
$\vec{S}_i +\vec{B}$, one expects that computing pulses inserted in
$0<t<T$ will be suppressed. In Fig. 6, we show that the distance of
$\rho_r$ to the desired reduced density matrix versus $\theta$ for
three different schemes. Indeed, the results clearly show that DD
pulses suppress the computing as well if one adopts the scheme of
computing while decoupling. Hence we shall adopt the scheme for
inserting computing pulses after decoupling pulses.

\begin{figure}[ht]
{\includegraphics[width=8cm]{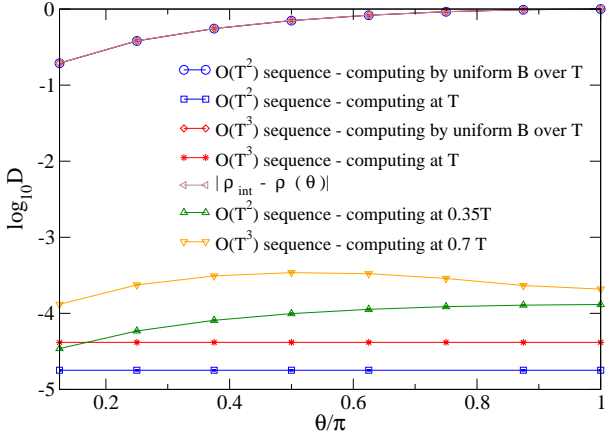}} \caption{{\protect\small
(Color online) Comparison of three differen schemes for inserting a
rotation $\theta$. Note that errors are smaller than sizes of figure
legends and $B=25$ Tesla so that $O(T^2)$ sequence outperforms
$O(T^3)$ sequence. Here initially the spin qubit is in the state
$|+\rangle$ and $\rho ({\theta}) $ is the corresponding density
matrix by applying a rotation of $\theta$ along $x$-axis on $|+
\rangle$ directly. $D = |\rho_r - \rho ({\theta})|$ is the distance
between the simulated reduced matrix and $\rho ({\theta})|$ with
$|A|=Tr \sqrt{A^{\dagger} A}$ }} \label{threescheme}
\end{figure}

In the following, we shall adopt the scheme that computing pulses
are inserted at $nT$ while the same DD pulses are applied between
computing pulses. We would like to demonstrate that in this scheme,
there will be cross-overs of the relative effectiveness for
different sequences. Based on the above optimized sequences within a
cycle of $T_c$, the performance at $nT_c$ can be deduced. Since
$U(nT_c) = U(T_c)^n$, we have
\begin{eqnarray}
  && U^{(N)}(nT_c) = \left\{ (-i)^p U^{(N)}_0 \left[ 1+ O(N \epsilon J \tau^2_p )\right]
  \right\}^n \nonumber \\
  && \approx \left\{ (-i)^p
  \left[ 1+ O(\epsilon J^{m-1} T^m)\right] \right\}^n \nonumber \\
  && \approx  \left\{ (-i)^p
  \right\}^n \left[ 1+ O(n\epsilon J^{m-1} T^m)\right].  \label{UTn}
\end{eqnarray}
Consequently $U(nT_c)$ is suppressed to $O(nT^m)$. To quantify the
performance at $nT_c$, it is useful to define the number of quantum
manipulation by defining $N_{op}$ as
\begin{equation}
L(N_{op}T_N) = \frac{1}{2} L(0). \label{nopL}
\end{equation}
Intuitively $N_{op}$ represents the maximal attainable number of
quantum operation before the qubit becomes decoherent. The exact
number of $N_{op}$ can be found by t-DMRG. In Fig. 7(a) we show the
effective dynamics for lower orders optimized sequences. It is clear
that decoupling pulses do suppress  the decoherence of a single
qubit. In Fig. 7(b), we plot $N_{op}$ versus $B$ for lower order
sequences. We find that $m=1$ sequence is the most optimized
sequence for low magnetic fields. This is consistent with estimation
in Fig.3.
\begin{figure}[ht]
{\includegraphics[width=9cm]{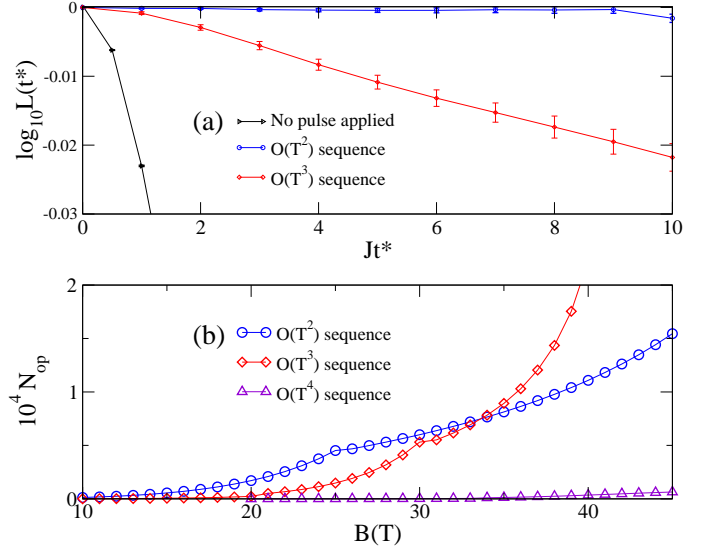}} \caption{{\protect\small
(Color online) (a) Effective dynamics of $L(t^*)$ for lower orders
optimized sequences shows better coherence than that of the original
dynamics. Here $J\delta t=0.01$, $\epsilon = -0.3$, and
$t^*=t/T_c(B)$ with $B=20T$. (b)$N_{op}$ for different orders versus
$B$ fields. In low fields, $O(T^2)$ sequence is the optimized. For
modest high fields, $O(T^3)$ sequence starts to win over $O(T^2)$
sequence.}} \label{HeisenbergSingleE}
\end{figure}

\section{Short Time and long time Dynamics of Two Qubits}

In this section, we extend the construction of pulse sequence to
multi-qubits. In particular, we shall examine the validity of our
construction for two qubits using t-DMRG. We start by considering
$n$ qubits denoted by $\vec{\sigma}^i$ with $i=1,2,\cdots,n$.
Following the derivation of Eqs.(\ref{UN1}) and
(\ref{timesequence}), one can move all the spin operators due to the
decoupling pulses to the left and obtain the corresponding $G^N(t)$.
The decoupling sequence generally introduces different history of
sign characterized by $f^i_\mu (t)$. We find $G^N(t) = \mathcal{T}
e^{ -i \frac{\epsilon}{2} \int_0^t ds \hat{R} (s)}$ with $ \hat{R}
(s)$ is given by
\begin{equation}
\hat{R} (s)= \sum_i \left[f^i_x(s) \sigma^i_x S^x_i(s) + f^i_y(s)
\sigma^i_y S^y_i (s) + f^i_z(s) \sigma^i_z S^z_i (s)\right].
\label{multitimesequence}
\end{equation}
If one assumes that different sequence is applied to different
qubit, it is clear that in addition to constraints set by
Eqs.(\ref{OT}), (\ref{OT2}), (\ref{OT3}) and (\ref{OT4}), there will be
extra constraints due to cross product of different qubit spins.
Therefore, it is clear that minimum number of pulses can be achieved
by setting all the pulse sequence the same: $f^i_\mu (t) \equiv f_\mu(t)$.
In this case, we find
\begin{equation}
  \hat{R} (s)= f_x(s) Q_x(s) + f_y(s) Q_y(s) +f_z(s) Q_z(s),
  \label{multitimesequence1}
\end{equation}
where $Q_\mu(s) =  \sum_i \sigma^i_\mu S^\mu_i (s)$.
In comparison to the case of a single qubit, here $Q_\mu$ replaces
the role of $\sigma_\mu S_\mu$. The commutators of $Q_\mu$ are given
\begin{equation}
  [Q_\mu(s), Q_\nu(s')]= \sum_{i,j}
  \sigma^i_\mu \sigma^{j}_\nu [S^\mu_i(s),S^\nu_j (s')].
  \label{commutator}
\end{equation}
According to Eq.(\ref{expansion1}), the $O(s^n)$ of
$S^\mu_i(s)$ is a n-fold commutator of $H_{bath}$ and
$S^\mu_i$. Since $H_{bath}$ only contains couplings between
nearest neighboring $S^\mu_i$, to $O(s^n)$, $S^\mu_i(s)$
contains spin operators up to $S^{\alpha}_{i \pm n}$. Therefore, to
$O(s^n{s'}^m)$ (i.e., to $O(T^{n+m+1})$), we find
$[S^\mu_i(s),S^\nu_j (s')]=0$ if $|i-j|>n+m$.
Consequently, commutators of $Q_\mu$ become
\begin{equation}
  [Q_\mu(s), Q_\nu(s')]= \sum_{i} \sigma^i_\mu
  \sigma^{i}_\nu [S^\mu_i(s),S^\nu_i (s')].
  \label{commutator1}
\end{equation}
It is then clear that for spin-$1/2$ qubits, because
$\sigma^2_\mu=1$, only commutators with $\mu \neq \nu$ contribute
dynamics. Furthermore, coefficients that are associated with these
commutators are exactly the same as those for a single qubit. Hence
Eqs.(\ref{OT}), (\ref{OT2}), (\ref{OT3}), and (\ref{OT4}) are also
the constraints for two qubits to suppress $U(T)$ to $O(T^4)$. In
other words, both the entanglement of two qubit and decoherence of a
single qubit can be optimized by the same sequence.

\begin{figure}[ht]
{\includegraphics[width=8cm]{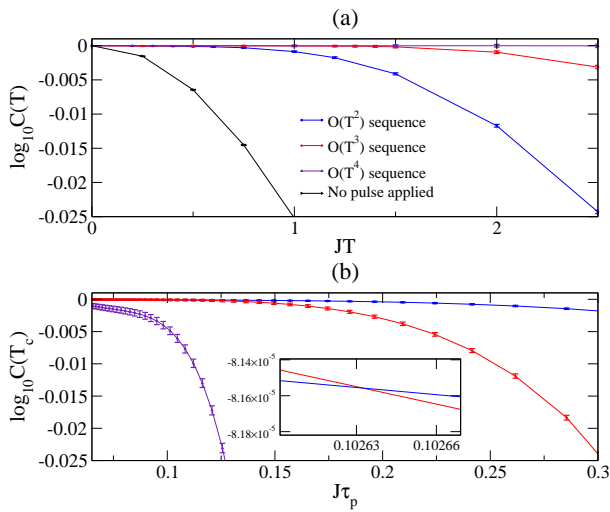}} \caption{{\protect\small
(Color online) (a) Concurrence $C(T)$ versus $T$ of different orders
simulated by t-DMRG for Heisenberg-like coupling with $\epsilon =
-0.3$. Clearly, $C(T)$ gets improved for increasing number of
orders. (b) $C(T_c)$ versus $J\tau_p$. It is clear that as $\tau_p$
decreases, $O(T^3)$ sequence starts to win over $O(T^2)$ sequence.}}
\label{HeisenbergSingleC}
\end{figure}
To check the validity of the above conclusion, we examine the
entanglement of two qubits. To characterize the entanglement, we
shall use concurrence as the measurement of
entanglement.\cite{Wootters} For a given reduced density matrix
$\rho(t)$, the concurrence is defined as
$C=\max\{\lambda_1-\lambda_2-\lambda_3-\lambda_4,0\}$, where
$\lambda_1\ge\lambda_2\ge\lambda_3\ge\lambda_4$ are the square roots
of the eigenvalues of the operator
$\rho(\sigma^y\otimes\sigma^y)\rho^*(\sigma^y\otimes\sigma^y)$ and
$\rho^*$ is the complex conjugation of $\rho$. In Fig. 8(a), we show
the concurrence $C(T)$ calculated by t-DMRG for various sequences.
It is seen that the sequence in Eq.(\ref{solution2}) indeed
outperforms other sequence. In Fig. 8(b) we show numerical results
of $C(T_c)$ versus $J\tau_p$ for different orders. It is clearly
seen that as $\tau_p$ decreases, $O(T^3)$ sequence starts to win
over $O(T^2)$ sequence. In Fig. 9(a), we show the effective dynamics
for the concurrence at $t^*=nT$. In comparison to the case without
decoupling pulses applied, it is clear that decoupling pulses do
improve the entanglement of two qubits. Fig. 9(b) shows that except
for $O(T^4)$ sequence with qubits at (39,40), decoupling pulses also
suppress the generation of entanglement. One of the reasons behind
the non-suppression of the entanglement generation for the $O(T^4)$
sequence is due to the large $T_c$ required by finite $\tau_p$.
However, as indicated by Eq.(\ref{commutator1}), the distance
between qubits is also an important factor. As a comparison,
$C(t^*)$ of the $O(T^4)$ sequence for two-qubits located at
different distances, (39,40) versus (30,50), are calculated. It is
seen that entanglement generation is suppressed only for qubits
located at (30,50), in agreement with conclusions based on
Eq.(\ref{commutator1}). In Fig. 9(c),we plot $N_{op}$ for different
orders versus $B$ fields. We find that at modest magnetic field, the
lowest order,$m=1$, is the most optimized sequence for preserving
the entanglement.
\begin{figure}[ht]
{\includegraphics[width=9cm]{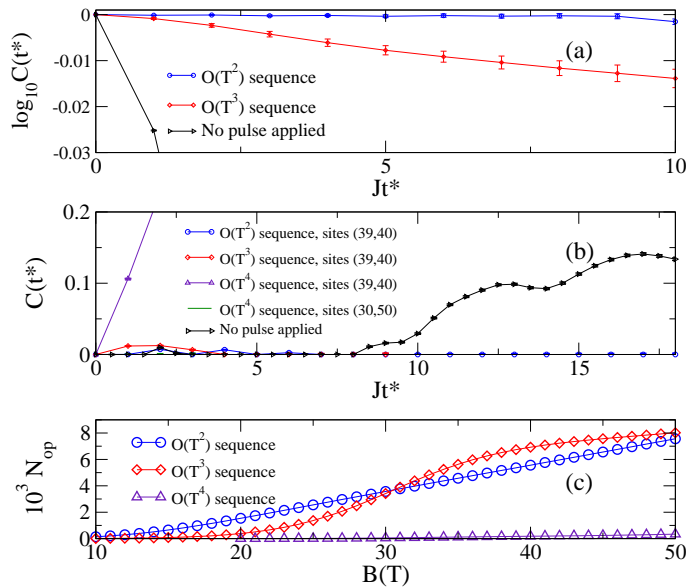}} \caption{{\protect\small
(Color online) (a)Effective dynamics of the concurrency. Here
$J\delta t=0.01$, $\epsilon = -0.3$, and $t^*=t/T_c(B)$ with
$B=20T$. (b)Effective dynamics for the generation of entanglement.
Here $J\delta t=0.02$, $\epsilon = -0.9$, and $t^*=t/T_c(B)$ with
$B=12T, 20T, 30T$ for $O(T^2), O(T^3)$, and $O(T^4)$ sequences
respectively. Note that errors are smaller than sizes of figure
legends. It is clear that entanglement generation is suppressed for
qubits that are far apart. (c)$N_{op}$ obtained from $C(t^*)$ for
different orders versus $B$ fields. In low fields, $O(T^2)$ sequence
is the optimized. For modest high fields, $O(T^3)$ sequence starts
to win over $O(T^2)$ sequence.}} \label{HeisenbergSingleCE}
\end{figure}

\section{Summary and Outlook}

In summary, feasibility of decoupling pulses that preserve the
coherence and entanglement of spin qubits with general couplings to
a quantum spin chain are examined. It is shown that error induced by
pulse-width accumulation restricts the consideration to sequences
with number of pulses being less than $10$ within each cycle. Within
each cycle $T$, exact nontrivial pulse sequences are constructed to
suppress the qubit-bath coupling to $O(T^{N+1})$ progressively with
number of pulses being $4,7,12$ for $N=1,2,3$. It is demonstrated
that computing after decoupling has the best performance. Therefore,
the performance is examined by counting number of computing pulses
that could be applied periodically at the end of each DD cycle.
Based on t-DMRG, our numerical results show that for modest magnetic
fields (10-40 Tesla) available in laboratories, the overall
performance is optimized when number of pulses in each cycle is 4 or
7 with pulse directions be alternating between x and z.

While so far our numerical results are obtained by using either a
single qubit or two qubits as demonstrations, our results also
provide insights for preserving coherence and entanglement of
multi-qubits. In fact, according to our analysis, in principle the
same sequences we obtained in this work can also dynamically
decouple multi-qubits from the environment in low magnetic fields.
For high magnetic fields, to obtain better coherence and
entanglement, one needs to suppress higher order terms. In this
case, however, one still needs to suppress lower orders. Therefore,
our results will still serve as a useful starting point for
sequences for higher magnetic fields.
\begin{acknowledgments}
We thank Profs. Hsiu-Hau Lin and Ming-Che Chang for useful
discussions. This work was supported by the National Science Council
of Taiwan.
\end{acknowledgments}

\appendix

\section{$O(T^3)$ Optimal Sequences}
In this appendix, we explicitly construct all possible sequences
that suppress $U(T)$ to $O(T^3)$. For this purpose, we first note
that operators appear in $U(T)$ in the order of $O(T)$ and $O(T^2)$
are $\sigma_x S^x$, $\sigma_y S^y$, $\sigma_z S^z$,
$[H_0,[H_0,S^x]]\sigma_x$, $[H_0,[H_0,S^y]]\sigma_y$, and
$[H_0,[H_0,S^z]]\sigma_z$. Requiring coefficients of these operators
to vanish yields Eqs.(\ref{OT}) and (\ref{OT2}). Eqs.(\ref{OT}) and
(\ref{OT2}) can be solved by using Mathematica to exhaust all
possible configurations of pulse directions. We find that in
addition to Eq.(\ref{OT3sequence}), the following generic sequences
are also
solutions \\
Sequence 1: pulse direction $xzxxzx$ \\
$\alpha_1=\alpha_2=\alpha_3=\alpha_5=\alpha_6=\alpha_7=1/8, \alpha_4=1/4$ \\
Sequence 2: pulse direction $xzxxyx$ \\
$\alpha_1=0.104715$, $\alpha_2=0.145282$, $\alpha_3=1/8$, $\alpha_4=1/4$ \\
$\alpha_5=1/8$, $\alpha_6=0.145282$, $\alpha_7=0.104715$ \\
Sequence 3: pulse direction $xzxzyz$ \\
$\alpha_1=0.0785$, $\alpha_2=0.1396$, $\alpha_3=0.1596$, $\alpha_4=1/4$ \\
$\alpha_5=0.1715$, $\alpha_6=0.0931$, $\alpha_7=0.1104$ \\
Sequence 4: pulse direction $xzxyzy$ \\
$\alpha_1=1/8$, $\alpha_2=0.095491$, $\alpha_3=0.1545$, $\alpha_4=1/4$ \\
$\alpha_5=0.1545$, $\alpha_6=0.095491$, $\alpha_7=1/8$. \\
In addition to the above sequences, equivalent sequences can be
formed by applying cycle permutations on $x$, $y$ and $z$.

\end{document}